\documentclass[12pt]{iopart}
 \usepackage{iopams}
 \usepackage{graphicx,color}
 \usepackage{multirow}

 \newcommand{\be}{\begin{equation}}
 \newcommand{\ee}{\end{equation}}
 \newcommand{\bea}{\begin{eqnarray}}
 \newcommand{\eea}{\end{eqnarray}}

\begin{document}

 \title{Radio data and synchrotron emission in consistent cosmic ray models}

 \author{Torsten Bringmann}
 \address{
 \mbox{II.}~Institute for Theoretical Physics, University of Hamburg, Luruper Chausse 149, D--22761 Hamburg, Germany}
 \ead{torsten.bringmann@desy.de}

 \author{Fiorenza Donato}
 \address{Dipartimento di Fisica Teorica, Universit\`a di Torino, 
 Istituto Nazionale di Fisica Nucleare, via P. Giuria 1, I--10125 Torino, Italy}
 \ead{donato@to.infn.it}

 \author{Roberto A. Lineros}
 \address{IFIC, CSIC-Universidad de Valencia, Ed.~Institutos,
 Apdo.~Correos 22085, E--46071 Valencia, Spain \& MultiDark fellow}
 \ead{rlineros@ific.uv.es}

 \begin{abstract}
 It is well established that phenomenological two-zone diffusion models of the galactic 
 halo can very well reproduce cosmic-ray nuclear data and the observed antiproton flux. 
 Here, we consider lepton propagation in such models and compute the expected galactic 
 population of electrons, as well as the diffuse synchrotron emission that results from their 
 interaction with galactic magnetic fields.  We find models in agreement not only with cosmic ray data but also
 with radio surveys at essentially all frequencies. Requiring such a globally
 consistent description strongly disfavors  very large ($L\gtrsim
 15\,$kpc) and, even stronger,  small ($L\lesssim 1\,$kpc) effective
 diffusive halo sizes. This has profound implications for, e.g.,
 indirect dark matter searches.
 \end{abstract}

 %%%%%%%%%%%%%%%%%%%%%%%%%%%%%%%%%%%%%
 %%%%%%%%%%%%%%%%%%%%%%%%%%%%%%%%%%%%%

 \section{Introduction}
 A wealth of observational data strongly suggests
  that diffusion governs the propagation 
 of galactic cosmic rays (CRs) \cite{Strong:2007nh}. 
 Any model for the underlying physical processes requires a basic assumption 
 about the geometry  of the region the CRs sweep. 
 In  diffusive models, the Galaxy is represented 
  by a thin disk sandwiched by a thick magnetic diffusive halo with cylindrical symmetry. 
 Given the intrinsic limitations to any such  model, as well as
 present-day CR data, a considerable uncertainty in the  propagation parameters  is generally unavoidable. 

 A realistic 3D modeling turns to extensive computer codes, such as Galprop 
 \cite{galprop,Moskalenko:2002yx}
 or Dragon \cite{dragon}, 
 aiming at a detailed description of sources, interstellar matter distribution,  
 magnetic field structure and diffusion phenomena.
  While reaching a high degree 
 of accuracy -- particularly needed for the gamma-ray component -- 
 such an approach is computationally expensive and does 
  not always make it straight-forward to extract physical answers  for the underlying processes and input parameters.  
  An effective 2D approach,
  on the other hand,
  %, based on a constant gas density in the disk, 
  benefits from analytical solutions to the spatial diffusion equation \cite{Maurin:2001sj,Maurin:2002hw,Donato:2001eq}, which 
   allows  fast computations and thus efficient scans
 of the  parameter space \cite{Putze:2010zn}. 
 Despite a small number of free parameters, it can consistently 
 describe both nuclear CRs \cite{Maurin:2001sj},  CR antiproton
   \cite{pbar} as well as electron and positron data 
 \cite{Delahaye:2008ua,Delahaye:2010ji}. 

 Here, we
 investigate  whether even radio data can be
 understood in this 2D framework. 
 We start by inferring  the galactic electron distribution from an 
 interpretation of large-scale radio survey data
 as synchrotron radiation and then
 compare this to the expectation in our propagation
 model.
 We find that radio observations  are indeed consistent
 with models that correctly describe CR data, providing
 a remarkable hint that our effective propagation model is not too far from
 a real picture of galactic phenomena, at least on kpc scales. 
 Furthermore, we show that our procedure can be used
 , in principle,
 as a new method to constrain propagation models which
 is complementary to using the boron over carbon (B/C) ratio \cite{Maurin:2001sj} or
  radioactive isotopes ratios (i.e.\ $^{10}{\rm Be}/^{9}{\rm Be}$) 
 \cite{Moskalenko:2002yx,Donato:2001eq,Putze:2010zn}. 

 The structure of this article is as follows.  We start by reviewing, in Section \ref{sec:synch}, how relativistic electrons produce synchrotron 
 radiation when propagating through  the galaxy. In Section \ref{sec:el}, we  compute the galactic electron distribution in our diffusion model and compare the expected synchrotron radiation in Section \ref{sec:data} to radio surveys at various frequencies, demonstrating that synchrotron radiation indeed provides a very promising means of both inferring  properties of the interstellar electron distribution and to provide constraints on the adopted  diffusion model. After a discussion of possible biases in our analysis in Section \ref{sec:disc}, we present our conclusions in Section \ref{sec:conc}.

 %%%%%%%%%%%%%%%%%%%%%%%%%%%%%%%%%%%%%
 %%%%%%%%%%%%%%%%%%%%%%%%%%%%%%%%%%%%%%%%%%

 \section{Synchrotron radiation}
 \label{sec:synch}
 Relativistic electrons\footnote{ Unless explicitly  stated otherwise, we will in the following use the term \emph{electron}
 to denote both electrons and positrons.  } emit synchrotron radiation while propagating through the galactic magnetic field 
 \cite{Ginzburg:1965su}. For electrons with energy $E$ and a magnetic field of strength $B$, the  emission power per unit 
 frequency is given by
 \begin{equation}
 \label{synch}
 \frac{dw}{d\nu} = \frac{\sqrt{3}\, e^3 B}{m_e c^2}\,
   \frac{2}{\pi} \int_{0}^{\pi/2} d\theta \ \sin\theta \ F\left(\frac{\nu}{\nu_c\sin\theta}\right)\,,
 \end{equation}
 where $\nu_c = {3 e B E^2}/{(4 \pi m_e^3 c^5)}$,  $F(x) = x \int_{x}^{\infty} d\zeta \ K_{5/3}(\zeta)$ and $K_{5/3}$ is a 
 modified Bessel function; we take the average over the angle $\theta$ between the electron momentum and $\vec B$ because we
 assume an isotropic electron distribution (note also that, in general, not only the regular but also the turbulent component
 of the galactic magnetic field contributes to the total signal -- which further motivates this average).

 For a given electron density $n_e(E,\mathbf{x})$, the expected intensity in synchrotron radiation is thus given by
 \be
 \label{intensity}
 I_\nu=\frac{1}{\Delta\Omega}\int_{\Delta\Omega}\!\!d\Omega\int d\ell\, J_\nu(\mathbf{x})\,e^{-\int_0^\ell d\ell'\,\alpha_\nu(\mathbf{x'})}\,,
 \ee 
 where $J_\nu=\int \frac{dn_e}{dE}\frac{dw}{d\nu}\,dE$ is the emissivity,  $\alpha_\nu$ the inverse of the absorption length and the integrations are taken along the line of sight towards the observed direction in the sky, averaged over an angular region $\Delta\Omega$. 
 Since radio emission is often associated with thermal phenomena, the intensity is traditionally stated in terms of temperature: with the Rayleigh-Jeans law in mind, $I_\nu=2\nu^2k_BT/c^2$, one can define the brightness temperature as
 \be
  \label{tb}
   T_b\equiv I_\nu c^2/(2\nu^2k_B)\,.
 \ee

 The galactic magnetic field has an average strength of  $\mathcal{O}(\mu{\rm G})$ \cite{microG}. 
 While its detailed structure in reality can be expected to be rather sophisticated  \cite{Bmodels},
 we will here adopt an effective approach and treat it to be spatially constant within the diffusion zone as we will only be interested in an average, large-scale description of  radio data above the galactic plane. In fact, this simplifying assumption seems necessary in order to be  consistent with the homogeneous diffusion coefficient that enters as a basic ingredient to our propagation model. We  verified that adopting instead a magnetic field falling off exponentially away from the disk --  in principle  also  consistent with the geometry of our diffusion model -- does not change significantly our predictions for the synchrotron flux (integrated along the line of sight) with respect to our choice of a constant magnetic field that abruptly vanishes at the border of the diffusion zone.
 This small difference can readily be understood  in terms of the propagation scale length, which is $\sim 300\,\rm{pc}$ for electrons with energies $\sim 3\,$GeV: 
 most of the electrons will simply remain relatively close to the galactic disk, where the magnetic field distributions are similar, thereby reducing the differences in the integrated synchrotron emission.

 For our analysis, we take into account absorption of synchrotron photons by both thermal electrons and ions from a possible hot gas component in the disk \cite{absorption}. Synchrotron self-absorption is another potential effect, but for $B\sim\mu$G it
  becomes numerically important only for frequencies below $1\,$MHz \cite{self_absorption}
  where it is however dominated by free-free absorption. 
 Finally, let us mention that for  an electron distribution following a power law, $dn_e/dE\propto E^{-\gamma}$, Eqs.~(\ref{synch}-\ref{tb}) tell us that also the synchrotron intensity, in the case of negligible absorption, follows a power law, $T_b\propto\nu^{-\alpha}$, with a spectral index
 \be
 \alpha=(\gamma+3)/2\,.
 \ee

 %%%%%%%%%%%%%%%%%%%%%%%%%%%%%%%%%%%%%
 %%%%%%%%%%%%%%%%%%%%%%%%%%%%%%%%%%%%%

 \section{Galactic electron population}
 \label{sec:el}

 \begin{table}[t]
 \centering
 \begin{tabular}{l||c|c|c||c|c}
      \multirow{2}{*}{\bf Model}  & \multicolumn{3}{c||}{\bf prop. parameters} & \multicolumn{2}{c}{ {\bf  radio data} ($\chi^2$/d.o.f.)}\\ 
      & $L\,[{\rm kpc}]$ & $K_0\left[\frac{{\rm kpc}^2}{\rm Myr}\right]$ & $\delta$ &  408\,MHz & 1.42\,GHz\\

 \hline
 \hline
  min & $1$ & $0.0016$ & 0.85 &  11.6 (6.8) & 11.9 (6.3) \\
  \hline
  med & $4$ & $0.0112$ & 0.70  & 4.9 (2.0) & 4.9 (2.0) \\
  \hline
  max & $15$ & $0.0765$ & 0.46  & 10.8 (4.8) & 8.9 (3.9) \\
 \end{tabular}
 \caption{\label{tab:models_res}  Benchmark models compatible with B/C data \cite{Donato:2003xg}. 
 Both 'min' and 'max' are clearly disfavored by radio data towards the galactic anti-center 
 (averaged over $10^\circ$ ($15^\circ$) and excluding the disk).
  }
 \end{table}
 The transport parameters  of the two-zone diffusion model introduced in Ref.~\cite{Maurin:2001sj} are determined from the
 B/C analysis  and correspond to the size of the diffusive halo of the Galaxy $L$,  the normalization of the diffusion
 coefficient $K_0$ and its slope $\delta$ (defined by $K=K_0 \beta R^\delta$, where $R=p/q$  is the rigidity), with a  rather
 strong degeneracy between the allowed parameters -- in particular between $K_0$ and $L$. 
 Diffusive reacceleration and convection, while crucial for the  nuclei analysis, have been shown to be only mildly relevant
 for  lepton fluxes  \cite{Delahaye:2008ua}. Indeed, the two processes shape GeV electron fluxes in the opposite direction,
 so that their combined effect is rather small; for that reason, and for the sake of simplicity, we do not take them into
 account here.  In this seminal analysis, we will for simplicity mainly refer to the three benchmark propagation models shown
 in Table \ref{tab:models_res}; among all models compatible with B/C data, these were shown to give the minimal, medium and
 maximal flux in antiprotons, respectively, that is expected from dark matter annihilations in the galactic halo
 \cite{Donato:2003xg}.

 High energy electrons are  produced  in 
 galactic accelerators such as supernova remnants (SNRs) or 
 pulsars  (primary electrons), as well as in hadronic interactions of galactic protons
  and helium nuclei with the  interstellar medium (secondary electrons). 
 We  calculate the primary $e^-$ flux from SNRs following Ref.~\cite{Delahaye:2010ji} 
 and the subdominant secondary $e^\pm$  component 
 as described in Ref.~\cite{Delahaye:2008ua},
  using a cylindrical gas density distribution in the disk as given in Ref.~\cite{Misiriotis:2006qq}.
 For the energy losses, we take into account 
 inverse Compton scattering off the interstellar radiation field, as well as synchrotron, 
 bremsstrahlung and ionization losses in the interstellar medium \cite{Putze:2010zn}.

 %%%%%%%%%%%%%%%%%%%%%%%%%%%%%%%%%%%%%
 %%%%%%%%%%%%%%%%%%%%%%%%%%%%%%%%%%%%%

 \section{Comparison to radio data}
 \label{sec:data}

 \begin{figure}[tbp]
 \centering
 \includegraphics[width=\columnwidth]{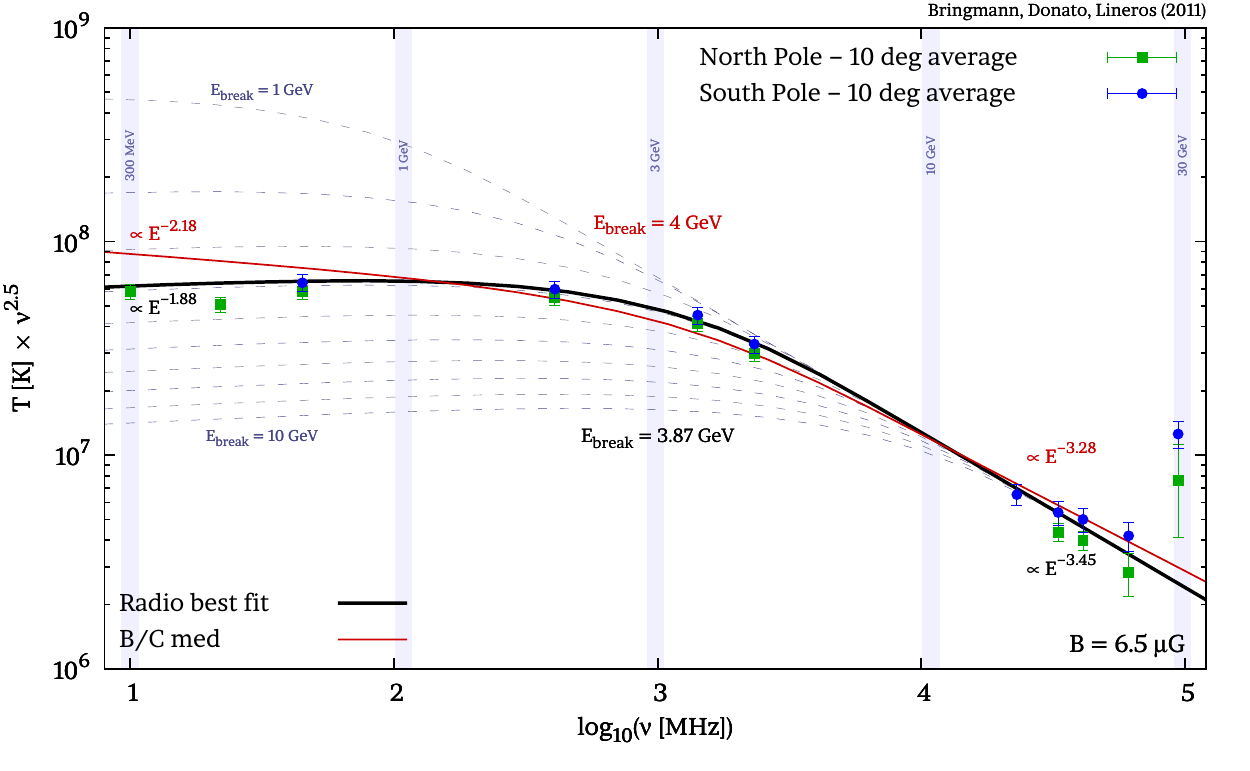}
 \caption{Observed radio fluxes vs. frequency.
 {Data points are generated using the GSM software~\cite{deOliveiraCosta:2008pb} which exactly reproduces the 
 observational data at 10~MHz~\cite{Caswell:1976}, 22~MHz~\cite{Roger:1999}, 45~MHz~\cite{Guzman:2010}, 
 408~MHz~\cite{haslam}, 1.42~GHz~\cite{Reich:1986}, 2.326~GHz~\cite{Jonas:1998}, as well as WMAP foregrounds at 
 23, 33, 41, 61 and 94~GHz~\cite{OliveraCosta:2006}.}
 The thick solid line represents the best fit for synchrotron radiation from an electron population 
 {(after propagation)} with $dn_e/dE\propto E^{-\gamma_1}$ ($dn_e/dE\propto E^{-\gamma_2}$) 
 for $E\!<\!E_{\rm br}$ ($E\!>\!E_{\rm br}$); the dashed lines indicate the effect of changing the best-fit 
 value $E_{\rm br}=3.9\,$GeV. The thin solid line shows a fit with a break in the energy losses for the 'med' model of Tab.~\ref{tab:models_res}; 
 see text for further details.}
 \label{fig:radio_freq}
 \end{figure}
 Starting from the 1960s, several groups have performed large-area radio surveys in the frequency range from about 1 MHz to
  100 GHz; for an extensive list, we refer the reader to Ref.~\cite{deOliveiraCosta:2008pb}. In this reference, the surveys
  at 0.010~\cite{Caswell:1976}, 0.022~\cite{Roger:1999}, 0.045~\cite{Guzman:2010}, 0.408~\cite{haslam},
  1.42~\cite{Reich:1986}, 2.326 GHZ~\cite{Jonas:1998}, as well as WMAP foregrounds at 23, 33, 41, 61 and 94
  GHz~\cite{OliveraCosta:2006}, were transformed to galactic coordinates and pixelized. In Fig.~\ref{fig:radio_freq}, we show
  the north and south pole at these frequencies, with the isotropic CMB component removed and averaged over a circular region
  with 10$^\circ$ in diameter.

 At these frequencies and latitudes, the dominant source of the radio signals should be synchrotron radiation (except for the
 excess seen in the 94 GHz band which is probably due to spinning dust \cite{deOliveiraCosta:2003az} and which we will not
 consider in the following). 
 {Free-free emission in general becomes important at frequencies $\nu\gtrsim1\,$GHz, but is
 physically subdominant when looking away from the galactic disk: adopting values for the thermal electron temperature and
 distribution as given in Ref.~\cite{arXiv:1108.6268}, we estimate the highest  free-free contribution (at 61\,GHz) to be
 less than 20\% in a cone towards the poles; given the error bars shown in  Fig.~\ref{fig:radio_freq}, this does not have a
 significant effect on our analysis.} 
 From the data, one can clearly distinguish two regimes with a different power-law
 behavior in frequency -- which directly translates to the necessity of a break in the spectral index of the 
 \emph{propagated} galactic electron population. If the propagated electrons are simply modeled with a broken power law, 
 the best-fit values for $dn_e/dE$ are a break in the spectral index $\Delta\gamma = 1.57^{+0.2}_{-0.25}$ at $E_{\rm br} =
 3.87^{+1.43}_{-1.17}\,$GeV, with $\gamma_2 = 3.45^{+0.15}_{-0.15}$ above $E_{\rm br}$ (note that the sharp break in the electrons gets smoothed because we
 use the full expression (\ref{synch}) for the synchrotron power rather than the often adopted monochromatic approximation). This result is fully
 consistent with, e.g.,  that of Strong et al.~\cite{andys_work}, see their Fig.~5, and 
 displayed   %in Fig.~\ref{fig:radio_freq} 
  as a thick, black solid line.  Here, we chose a fiducial value of
 $B=6.5\,\mu$G; a different value would simply change the  \emph{location} of  $E_{\rm br}$ and the (arbitrary) overall
 normalization (by a factor $\propto B^2$), but not the functional dependence on $\nu$ (we take the opportunity to remind the
 reader that $B=6.5\,\mu$G is an \emph{effective} value which takes into account both the average regular \emph{and}
 turbulent component of the galactic magnetic field).

 \begin{figure}[tbp]
 \centering
 \includegraphics[width=\columnwidth]{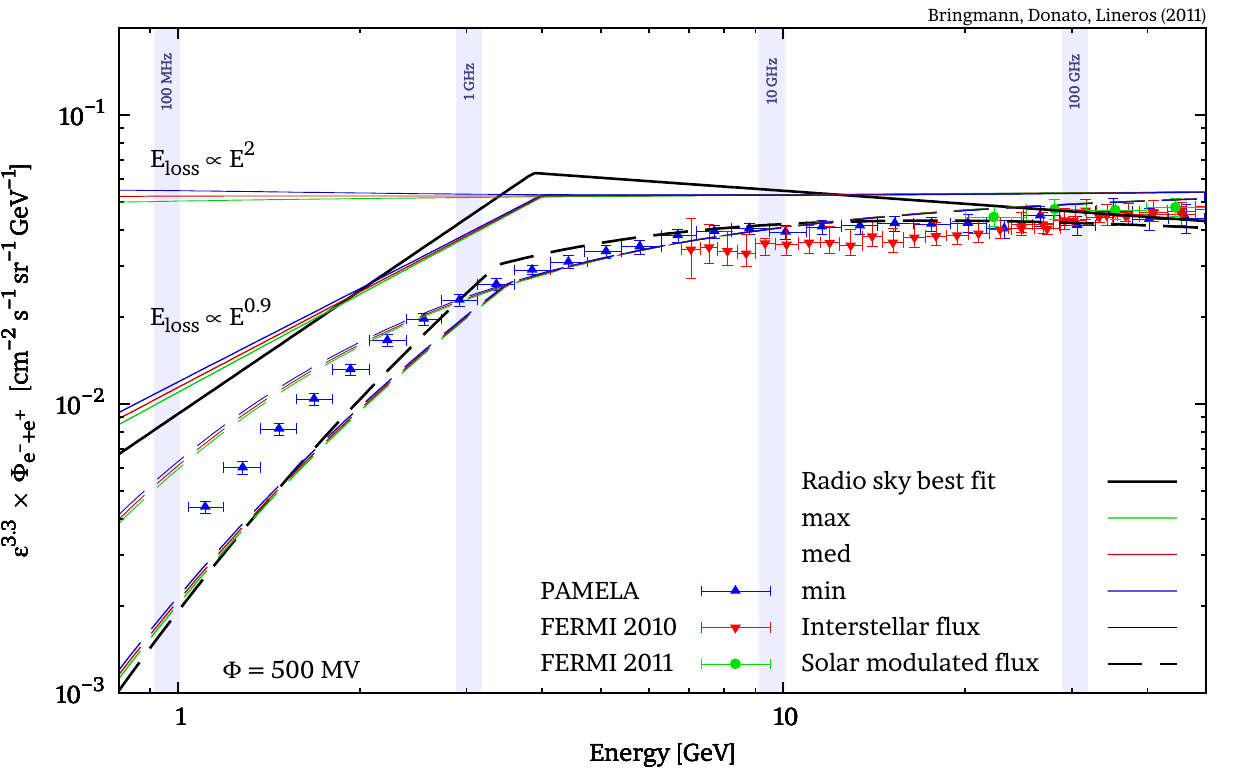}
 \caption{Measured  spectrum in CR electrons ($e^-$ for the case of PAMELA) 
 below 50\,GeV \cite{electron_pamela_2011,electron_fermi_2010,electron_fermi_2011}
  and  predictions according to Refs.~\cite{Delahaye:2008ua,Delahaye:2010ji}, assuming a spectral break
 in the energy losses. This spectrum is consistent with 
  the radio data shown in Fig.~\ref{fig:radio_freq} 
  (vertical bands indicate the corresponding frequency of  synchrotron radiation).}
 \label{fig:leptons}
 \end{figure}
 
%%%%%%%%%%%%%%%%%%%%%%%%%%%%%%%%%%%%%%%%%%%%%%%55
 Such an electron distribution is actually also in rather good agreement 
 with the observed CR electron  fluxes (note that only electrons well below 100 GeV are relevant 
 to our discussion because the spectrum at higher energies is likely to be
 dominated by local sources).  
 In Fig. \ref{fig:leptons}, we show the Pamela $e^-$ data \cite{electron_pamela_2011} together with the
 $e=e^++e^-$ data taken by Fermi-LAT  \cite{electron_fermi_2010}. Very recently,  the Fermi-LAT Collaboration has performed a separation
 between  observed positrons and electrons above 20 GeV using the Earth magnetic field  as a spectrometer \cite{electron_fermi_2011},
 confirming their previous  results on the total lepton fluxes. 

Along with the data, we plot in Fig. \ref{fig:leptons} theoretical predictions for the interstellar and solar modulated fluxes.
 Black lines correspond to the radio best fit (see Fig.~\ref{fig:radio_freq}), 
 where electrons are naively shaped by two power laws with break of  $\Delta\gamma=1.57$.
 We also show the resulting flux at Earth after propagation according to
 Refs.~\cite{Delahaye:2008ua,Delahaye:2010ji} and for the propagation models listed in 
 Table~\ref{tab:models_res}; here, we included for comparison a break of 1.1 in the energy loss 
 term, making the hypothesis that losses effectively follow $E^{0.9}$ below $E_{break}$ (for the 'med' model, this is in addition shown as a red solid line in Fig.~\ref{fig:radio_freq}). 
 In order to obtain the curves shown in Fig. \ref{fig:leptons}, we choose  the normalization and spectral index $\gamma_{\rm inj}$ of the
 primary $e^-$ spectrum in such a way as to fit the \emph{total} (primary and secondary) $e^-$ spectrum  to the low-energy PAMELA data (we checked that
 adding  the Fermi electron data, given their energy range and  error bars,   would not  significantly affect this normalization). We
 stress that our normalization is  consistent with Fig.~\ref{fig:radio_freq}, i.e.~it reproduces both the the radio data (with
 $B=6.5\,\mu$G as stated above) and, roughly, the low-energy lepton data  \cite{electron_pamela_2011}.
 
  For energy losses that do not significantly change during 
 the typical path an electron propagates, 
 the spectral index of the expected electron flux is approimately given by \cite{Delahaye:2010ji}
 \be
  \gamma\approx\gamma_{\rm inj}+\frac{1}{2}(\beta+\delta-1), 
 \label{eq:gamma_tilde}
 \ee
 where $\gamma_{\rm inj}$ is the spectral index of the injected electrons (before propagation) 
 and the energy loss term is assumed to scale like $b\equiv-\partial_t E\propto E^\beta$. 
 This relation is not more valid, however, if electrons experience different energy 
 loss regimes while wandering through the Galaxy, depending for example on the 
 time spent in the halo (which is gas free) or in the disk. If energy losses are dominated by 
 a term specific to the local environment where the electron spends 
 most of its time, $b\propto E^\beta_{\rm loc}$, one can derive from the general expressions given 
 in Ref.~\cite{Delahaye:2010ji} that the expected relation for the observed spectrum
instead becomes
 \be
  \gamma\approx\gamma_{\rm inj}+\frac{1}{2}(-\beta+\delta-1) + \beta_{\rm loc}\,,
 \label{eq:gamma_tilde_loc}
 \ee
 which reduces to Eq.~(\ref{eq:gamma_tilde}) for $\beta_{\rm loc}=\beta$.
We note therefore that 
the break illustrated in Figs.~\ref{fig:radio_freq} and \ref{fig:leptons}, taking into account the viable scatter in $\Delta\gamma$,  might simply reflect the fact that at lower 
 energies the scattering on thermal ions and electrons ($b_{\rm loss}\propto E$) becomes more important than inverse Compton losses
 ($b_{\rm loss}\propto E^2$). This interpretation is also consistent with the \emph{location} of the break that is determined by the
 relative strength of these processes: taking into account uncertainties in the radiation, magnetic field and gas densities, we
 expect the transition to occur in the range $1\,$GeV$\lesssim E_{\rm break}\lesssim10\,$GeV. In fact, as also shown in
 Figs.~\ref{fig:radio_freq} and \ref{fig:leptons}, a corresponding value of $\Delta\beta_{\rm loc}=1.1$ (which we choose slightly larger than
 1 in order to capture possible effects of reacceleration) does show a very reasonable agreement with both radio and lepton data.

\begin{figure}[tbp]
 \centering
 \includegraphics[width=\columnwidth]{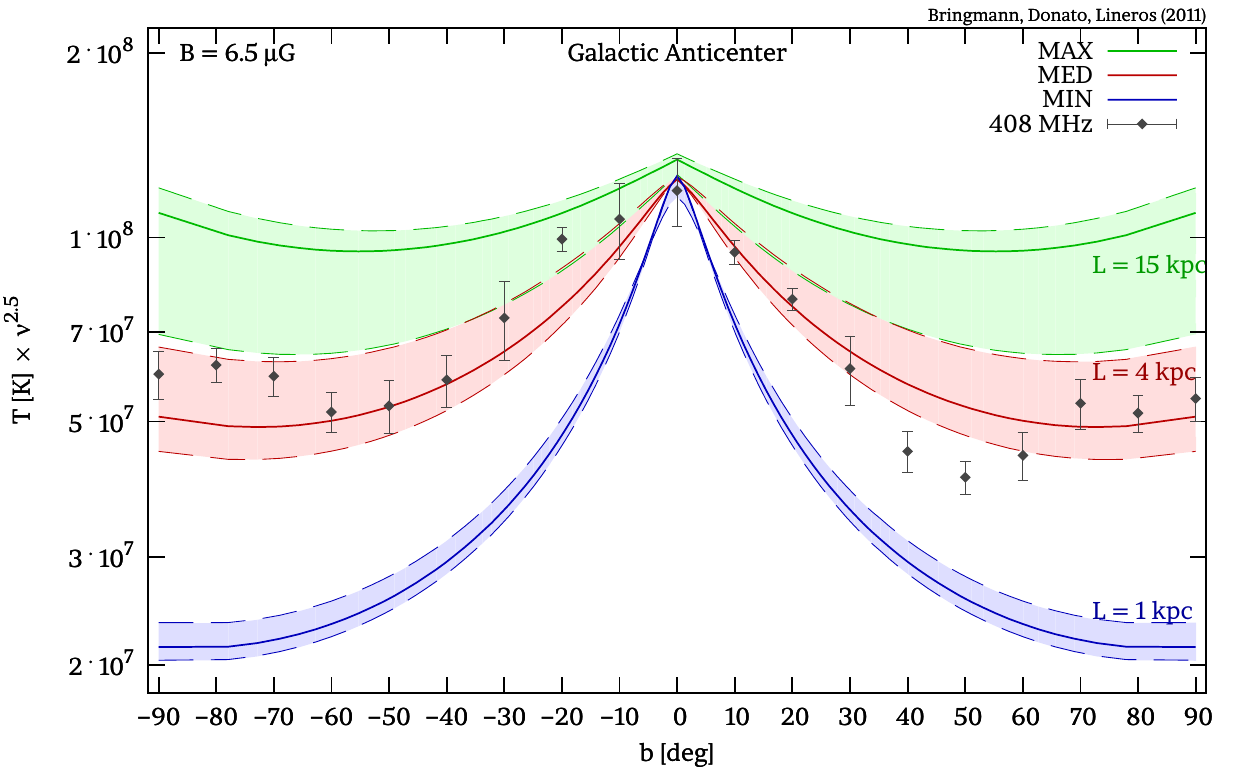}
 \caption{Brightness temperature versus latitude at 408 MHz. Solid lines correspond to the propagation models given in Table \ref{tab:models_res}. The shaded areas are obtained by varying $K_0$ and $\delta$, at fixed $L$, over the full range compatible with B/C data.}
 \label{fig:radio_lat}
 \end{figure}

%%%%%%%%%%%%%%%%%%%%%%%%%%%%%%%%%%%%%%%%%%%%%%%%%%%%%%%%%%%%%%%%%%%%%%%%%%%%%%%
 Let us now turn to the \emph{angular} shape of the radio signals, which we show in Fig.~\ref{fig:radio_lat} for the case of the Haslam \cite{haslam} map at 408 MHz. In the same figure, we indicate the expectation for the models of Tab.~\ref{tab:models_res}, including the full variation of the parameters $K_0$ and $\delta$ within the range compatible with the B/C analysis performed in Ref.~\cite{Maurin:2001sj}. One can clearly see that the size of the diffusive halo has a rather strong impact on the angular shape of the resulting synchrotron signal. Demanding consistency with B/C data, small halo sizes $L\sim1\,$kpc are essentially excluded, but also large values $L\gtrsim15\,$kpc show some tension with radio data. Note that the magnetic field normalization
 does not affect  the angular shape of the synchrotron emission; here, it
  was chosen such as to be consistent with Figs.~\ref{fig:radio_freq} and \ref{fig:leptons}.

 In order to demonstrate the constraining power of the angular shape 
 of  radio data 
 for diffusion models, we next
 treated the overall normalization of the synchrotron signal (aka the magnetic field) as a free parameter to minimize $\chi^2$ for the min/med/max models. The result is shown in Tab.~\ref{tab:models_res} for the  408\,MHz and 1.42\,GHz maps;  as anticipated, it is very difficult to reconcile  radio data with the max and, even more so, the min  model. Similar conclusions apply to all  surveys with $\nu<408$\,MHz, while for $\nu>1.42$\,GHz  the  data generally start to reproduce the expected synchrotron pattern  worse, even for the med model; 
 this is likely due to molecular clouds and pulsars (and/or SNRs), mostly in the southern hemisphere, that produce bremsstrahlung emission.
 We note that even the overall fit quality, for the min model, is quite good for angular averages of at least $15^\circ$; on smaller scales, on the other hand, we can obviously  not expect our effective model to  reproduce the detailed features that are visible in the radio sky.
 While these results already indicate the power of the method and thus warrant a  more detailed analysis of the propagation parameter space and the magnetic field structure \cite{ourpaper}, this  is beyond the scope of the present work.

 %%%%%%%%%%%%%%%%%%%%%%%%%%%%%%%%%%%%%
 %%%%%%%%%%%%%%%%%%%%%%%%%%%%%%%%%%%%%
 \section{Discussion}
 \label{sec:disc}

 Let us now discuss which assumptions in the analysis presented above are  affected
 by uncertainties that might bias our conclusions. We have explicitly verified that
 the absorption of synchrotron radiation by  thermal electrons
 is negligible for $\nu \gtrsim $ 10\,MHz and
 insensitive to the thermal electron
 temperature (for 4000\,K$<T_e<$8000\,K) below this frequency;
 for very low values ($T_e\lesssim$3000\,K)  absorption modifies 
 the spectral shape in a way  strongly disfavored by the data.
 Weak uncertainties  are derived from the ensuing
 bremsstrahlung radiation at high frequencies, whose
 contribution emerges around the GHz and gets stronger with
 lower $T_e$.

 We checked that a break in the diffusion coefficient does not induce
 relevant changes in the frequency spectrum, in contrast to the
 break in the (total) propagated spectrum discussed above;
  we found that a break in the source spectrum of primary electrons alone, on the other hand, would result in an overproduction of synchrotron radiation from secondary electrons at frequencies $\nu\lesssim100\,$MHz.
 We  also verified that modifying $K_0$ or $L$ merely shifts
 the spectrum $w.r.t.$ $\nu$ by  changing its normalization;
 as a function of $\nu$, the three benchmark models
 of Tab.~\ref{tab:models_res}, e.g., simply scale up the spectrum by a factor of 5 when going from min to max. 
 A degeneracy in  $B$, $K_0$ and $L$  (marginally also $\delta$)
 becomes manifest  even when looking at the
 radio spectrum as a function of the latitude.
 It is only when we turn to physical (i.e.~B/C  compatible) values
 of the propagation parameters that the degeneracy is
 partially broken in the spectrum $w.r.t.$ the
 latitude, see Fig. \ref{fig:radio_lat}.

 %%%%%%%%%%%%%%%%%%%%%%%%%%%%%%%%%%%%%
 %%%%%%%%%%%%%%%%%%%%%%%%%%%%%%%%%%%%%
 \section{Conclusions and Outlook}
 \label{sec:conc}

 The connection between synchrotron radiation and radio data
 provides a very interesting means of inferring properties of the interstellar galactic electron population at low energies   \cite{old_refs}.
 We have shown that  the radio sky from MHz to GHz, when averaged over large scales,
  can be understood in terms of
 synchrotron emission of diffused galactic $e^\pm$
 predicted in a model consistent with many other cosmic observables,
  including CR nuclei
 (from protons to iron), radioactive isotopes and antiprotons. 
 This is a rather new result and strong evidence that we are not too far
 from having a global picture of the phenomena occurring in the galaxy.

 We also found first indications that such a 
 description breaks some of the degeneracies encountered when constraining the properties of the diffusive halo with nuclear CR data.
 A more detailed analysis of the allowed space of propagation parameters  will be treated in a forthcoming publication \cite{ourpaper}.  However, let us stress that we have already presented preliminary evidence for $1\,{\rm kpc}\lesssim L\lesssim15\,{\rm kpc}$. We note that  in particular the $\bar p$ flux from dark matter annihilations in the galactic halo  
 is mostly sensitive to the volume probed by the magnetic diffusion zone; a lower bound on $L$ will thus have important implications for indirect dark matter searches.

 %%%%%%%%%%%%%%%%%%%%%%%%%%%%%%%%%%%%%
 %%%%%%%%%%%%%%%%%%%%%%%%%%%%%%%%%%%%%
 \ack
 We thank A.~Strong and M.~Regis for valuable discussions and acknowledge use of the HealPix software \cite{healpix}. We also would like to thank the anonymous referee for useful comments that helped to improve this article. TB was supported by the German Research Foundation (DFG) through  Emmy Noether grant BR 3954/1-1.
 FD and RL thank the LEXI initiative and the theoretical astroparticle group at the university of Hamburg for financial support and hospitality.  
 RL was supported by the EC contract UNILHC PITN-GA-2009-237920, by the Spanish grants FPA2008-00319, FPA2011-22975, MultiDark CSD2009-00064 (MICINN), and PROMETEO/2009/091 (Generalitat Valenciana).

 \begin{footnotesize}
 {\it Note added.} 
 In the final stage of preparing this manuscript, we became aware of Refs.~\cite{Jaffe:2011qw,andys_work} also investigating  the complementary information from CR and synchrotron data on galactic electrons 
 %by using the GALPROP code
 (with, however, a somewhat different focus compared to ours). 
 \end{footnotesize}

 %\bibliography{bib/radio_bib} % this is the reference to bibtex file!

 \end{document}